\documentstyle[aps,multicol,psfig,amssymb,graphicx,exscale,
fontenc,latexsym]{revtex}

\def \FIGUREWIDTH{8.2cm}

\def  \Lk{ {\em Lk} }
\def  \Wr{ {\em Wr} }

\def \beq{\begin{equation}}
\def \eeq{\end{equation}}

\begin{document}

\psrotatefirst

\draft 
\title{Topological and geometrical entanglement in a model of circular 
DNA undergoing denaturation.}

\author{Marco Baiesi,$^{1,}$\footnote{E-mail: baiesi@pd.infm.it} 
Enzo Orlandini,$^{1,}$\footnote{E-mail: orlandini@pd.infm.it} 
and Attilio L.~Stella$^{1,2,}$\footnote{E-mail: stella@pd.infm.it} }
\address{$^1$ INFM-Dipartimento di Fisica,\\
Universit\`a di Padova, I-35131 Padova, Italy\\
$^2$ Sezione INFN, Universit\`a di Padova, I-35131 Padova, Italy }
\date{\today}

\maketitle

\begin{abstract}
The linking number (topological entanglement) and the
writhe (geometrical entanglement) of a model of circular
double stranded DNA undergoing a thermal denaturation transition
are investigated by Monte Carlo simulations.
By allowing the linking number to fluctuate freely in equilibrium  we see
that the linking probability undergoes an abrupt variation (first-order)
at the denaturation transition, and stays close to 1
 in the whole native phase.
The average linking number is almost zero in the denatured phase
and  grows  as the square root of the chain length, $N$, 
in the native phase.
The writhe of the two strands grows as $\sqrt{N}$ in both phases.
\end{abstract}

\begin{multicols}{2} \narrowtext

\section{Introduction}
\label{sec:intro}

It is well known  that
topological (knotting and linking) and geometrical (supercoiling) 
entanglements in long DNA molecules are critical to the functioning of the
cell \cite{SKI87}. For this reason 
 there exist enzymes such as topoisomerases and 
recombinases which facilitate cellular metabolism by changing the 
geometry and the topology of DNA\cite {DC,Wang1,WC,Wang2}.

For closed loops of double-stranded DNA (dsDNA) the consequences
of the double-helix linking number $\Lk$ being non zero have been extensively
studied \cite{W69,F71,C90,VC94}. For example, it is known that a large imposed 
linking number leads the chain to twist upon itself. This is the phenomenon 
of {\sl supercoiling} in DNA, analogous to the familiar buckling of a 
twisted tube or wire\cite{VLRWL65}. 
The linking entanglement is believed to play 
an important role in  
structural transitions of double stranded chains such as 
local denaturation and cruciform structures formation.
In particular at the melting transition it is possible that progressive 
supercoiling of the native part 
of the molecule affects the denaturation process and there are indeed
experimental indications that
supercoiled structures denature
at higher temperature and over a broader temperature range than   
DNA molecules in the relaxed state\cite{BB78,GBL81}.
However, despite this evidence of the effects that topology 
and geometry can have on the denaturation transitions of circular dsDNA's,
theoretical studies of thermal denaturation have in general
neglected this aspect up to know. 
Indeed, starting with the seminal work of Poland and Sheraga\cite{PS66}, 
statistical mechanics studies of denaturation  have focussed mainly on
the nature of the  transition  and on how this depends
on properties such as the self  and mutual avoidance  of the strands 
\cite{CCG00,KMP00,COS02}, 
the bending rigidity of the bound segments\cite{TDP00}, 
or the inhomogeneity of the base pair sequence\cite{CH97}.
An attempt to include topological constraints in statistical models
of thermal denaturation has been made 
recently by Rudnick and Bruinsma\cite{RB01}.
 This study, carried on within the Poland and Sheraga model, is based on
the introduction at mean field level of an
elastic strain energy concentrated
in the native part of the chain. Their model completely neglects the 
$3D$ nature of the entanglement complexity of the double stranded chain.
In order to reach more firm conclusions about the effects of the torsional 
strain on the nature of thermal denaturation, one needs of course models in 
which these effects coexist with a realistic representation of the 
geometrical and topological entanglement of DNA.

In the present article we address  the problem of 
describing properly the topological and geometrical complexity of DNA when it 
undergoes the denaturation transition. More precisely, we set up a model in 
which, while denaturation is induced by energetic 
factors not directly related to the entanglement complexity of the chain, 
the topological and geometrical properties can be meaningfully represented 
and monitored in equilibrium. This approach, which has some complementarity
with respect to that of Ref.~\cite{RB01}, should be seen as a first step 
towards descriptions that embody also torsional strain interactions 
induced by a linking constraint, which are not considered here.

\section{Model and phase diagram}
\label{sec:simu}

We consider a model
of circular double stranded DNA where 
the two strands are 
mimicked by two non-overlapping self-avoiding polygons (SAP) of $N$ edges
on the cubic lattice.
Suitable short range attractive interactions between homologous pairs
of monomers of the two strands  are then considered to induce the formation
of the double structure.
 Each SAP is rooted at one vertex
and the  rooted vertices of the two SAP's are fixed to remain always a 
lattice distance apart. 
We identify the configuration $\omega$ of the two SAP's 
by the positions  $\vec{r}_1(i)$ and $\vec{r}_2(i)$, $i=1,2,\ldots N$, of 
the visited lattice vertices.  A binding energy $-\epsilon$, ($\epsilon>0$)
is associated to each  pair of vertices
$(\vec{r}_1(i),\vec{r}_2(j)) $ having $i=j$ and distance
$| \vec{r}_1(i)-\vec{r}_2(j) | \le \sqrt{3}$  in lattice units.
We call such kind of pair a {\sl contact}. The extension of the
range of the interaction to the third neighbors 
is suggested by the need of conferring a reasonable degree of flexibility
to the two SAP's when they are bound together to form the double chain 
structure. 
At low temperature, 
shorter ranges of interaction would indeed cause 
double stranded (ribbon-like) structures that are quite rigid
\cite{OBMSM02} and, even if rigidity effects are not expected to be relevant 
for the asymptotic features of the denaturation transition \cite{COS02}, 
their reduction can avoid slow crossover effects.

The Hamiltonian of a dsDNA configuration $\omega$ is
\beq
H(\omega) \equiv - m(\omega) 
\ ,
\label{Ham}
\eeq
where $m(\omega)$ is the number of contacts as defined above.
The canonical partition function $Z$  at an inverse
temperature $\beta=1/T$ is
\beq
Z \equiv \sum_{\omega} e^{-\beta H(\omega)}
\label{Z}
\eeq
and most thermodynamic quantities follow as weighted averages
normalized by $Z$.
The binding energy is taken to be the same all along the strand, i.e.,
we neglect the heterogeneity of base pair interactions of specific
sequences. This is a reasonable first approximation if we consider that a
single unit lattice step should correspond to a persistence length of the 
strands. So $\epsilon$ represents in fact an average binding energy for a 
set of several base pairs.  At $\beta=0$ the two strands behave as independent SAP's, 
whereas at sufficiently high values of $\beta$ the two SAP's are  
bound, i.e a macroscopic number ($\propto N $) of contacts occurs.
In analogy with similar previously studied models\cite{CCG00,OBMSM02},
a melting transition is expected to occur at some $\beta=\beta_c$.
Coming from low temperatures, this transition is driven by the formation of denatured loops, whose 
length can be measured by the number of unbound monomers $\ell$ comprised 
between two consecutive contacts of the
corresponding strand segments \cite{KMP00,COS02}.

\begin{figure}[!tbp]
\centerline{
\psfig{file=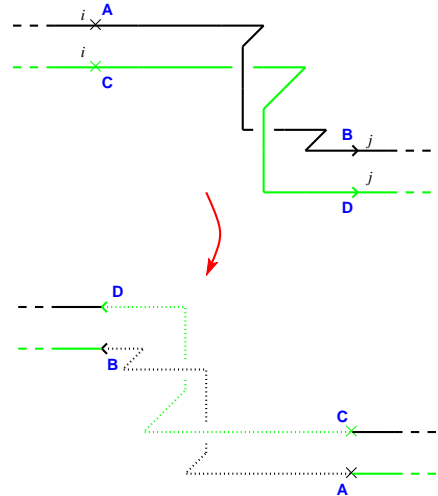,width=6cm}}
\vskip 0.2truecm
\caption{Schematic representation of the double-inversion move:
two  monomers $i$ and $j$ are selected randomly (with $i<j$) 
along the SAP sequence and a move is tried if
 the positions $\vec{r}_1(i)$ and $\vec{r}_1(j)$ 
(first SAP) and $\vec{r}_2(i)$ and $\vec{r}_2(j)$ (second SAP) 
satisfy the relation
$\vec{r}_1(i)-\vec{r}_2(i) = \vec{r}_1(j) - \vec{r}_2(j)$.
The proposed move
is an inversion of the portion of the two chains between $i$ and $j$ 
 with respect to their middle point
$\vec{r}_m = 1/2(\vec{r}_1(i) + \vec{r}_2(j))$.
 }
\label{fig:01}
\end{figure}

Configurations have been sampled by Monte Carlo (MC) methods.
Since the system is strongly interacting we have adopted 
a multiple Markov chain approach in which one samples 
simultaneously at a variety of different temperatures
and ``swaps'' configurations between contiguous temperatures.
The swap probability is chosen so that the limit distribution of the
process is the product of the Boltzmann distributions at the individual 
temperatures \cite{TJOW96}. 
The underlying (symmetric) Markov chain used  is based on a combination of 
pivot moves for SAP \cite{MOS90} and of local moves \cite{VS61}. In addition, 
to increase the mobility of the MC sampling in the native phase, we have 
introduced a  new, ``double inversion'' move (see Fig.~\ref{fig:01}). 
This move has the 
advantage that it may change the mutual entanglement of the two SAP's 
while keeping  the number of contacts fixed; it is thus particularly 
effective in sampling  configurations in which the two strands 
are strongly bound and linked. 
The result of applying this move is a net  increase of the mobility 
of the MC in the low temperature, native phase.

In our simulations, each MC step
consists of $O(1)$ pivot and double inversion moves and $O(N)$ local moves.
By running in parallel  a number $M\approx 30$ of  Markov
chains we are able to obtain good sampling for chains up to 
$2 N=1200$, from $\beta=0$ up to $\beta=1.1$.
For all lengths we  obtain samples with at least  $10^{5}$ 
independent data points
 for almost all $\beta$'s considered. Only for the highest
$\beta$'s the correlation of successive samples reduces 
the number of independent data points to $\sim 10^{4}$. 
To interpolate the obtained data at intermediate temperatures we have 
used the multiple histogram method \cite{mhm}.

\begin{figure}[!bt]
\vskip 0.3truecm
\centerline{
\psfig{file=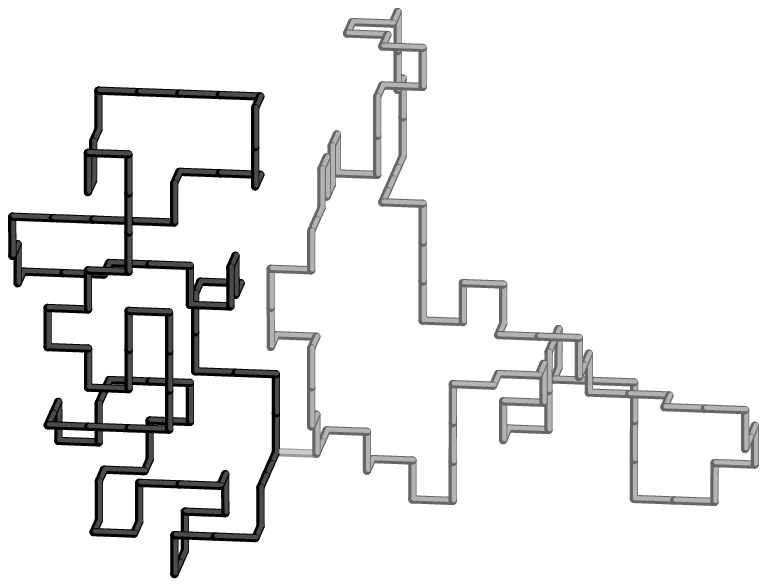,angle=270,width=6.3cm}}
\vskip 0.7truecm
\centerline{
\psfig{file=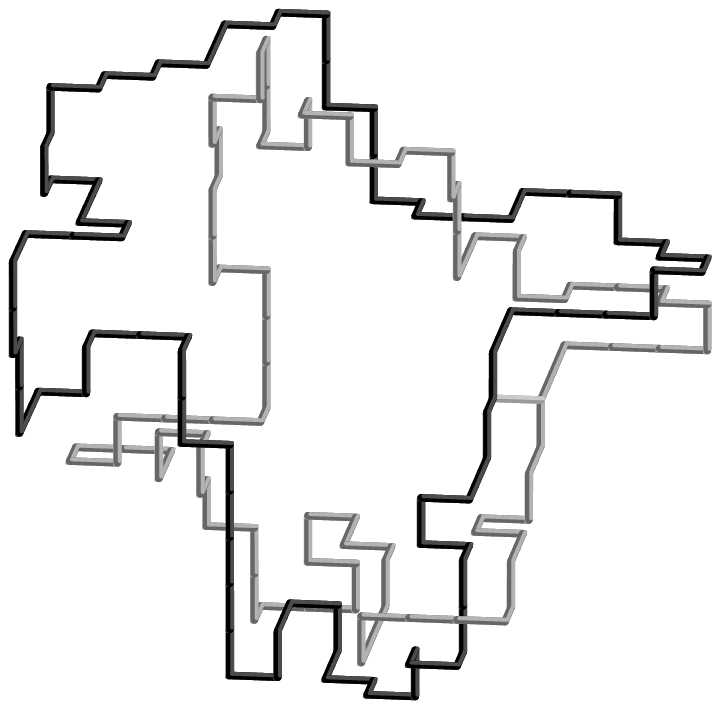,angle=270,width=6cm}}
\vskip 0.7truecm
\caption{Snapshots of typical configurations at $\beta$ lower (top)
and higher (bottom) than the denaturation inverse temperature $\beta_c$.
}
\label{fig:02}
\end{figure}

In order to study the entanglement of the configurations as a function
of temperature and in particular at the melting transition, we need
first to characterize the phase diagram of the model and
to estimate $\beta_c$.
Two regimes are manifestly present: a low $\beta$ regime characterized by
two unbound and independent SAP's,  and a high $\beta$
phase dominated by configurations in which the two strands are
strongly bound (see Fig.~\ref{fig:02}).

\begin{figure}[!bt]
\centerline{
\psfig{file=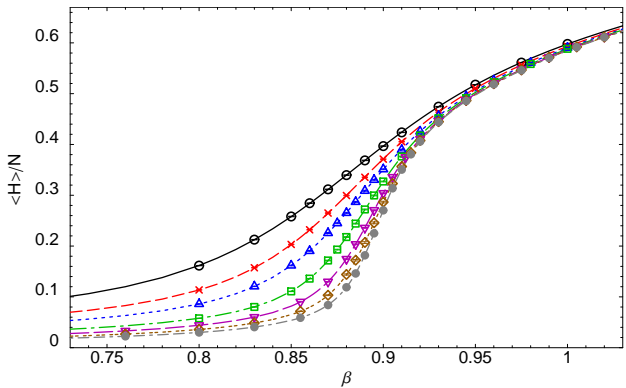,angle=0,width=\FIGUREWIDTH}
}
\vskip 0.1truecm
\caption{Plot of the energy density $\langle H \rangle/N$ 
as a function of $\beta$, for $N = 100 (\circ)$, $150 (\times)$,
 $200 (\vartriangle)$,  $300 (\Box)$, $400 (\triangledown)$, 
$500 (\diamond)$, $600 (\bullet)$.
As $N$ increases, two behaviors are clearly distinguished: 
for $\beta \lesssim0.9$, $\langle H \rangle/N\to 0$, 
while $\langle H \rangle/N\to$const for $\beta\gtrsim0.9$.
}
\label{fig:03}
\end{figure}

A transition between these two regimes is suggested by the behavior
of the average energy per monomer 
\beq
\frac{\langle H \rangle}{N} =
\frac{1}{Z N} \sum_{\omega} H(w) e^{-\beta H(\omega)}
\label{U_Ndef}
\eeq
as a function of $\beta$ (see Fig.~\ref{fig:03}).

\begin{figure}[!bt]
\centerline{
\psfig{file=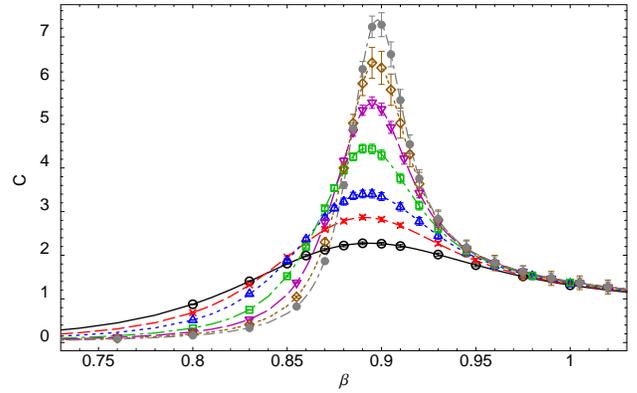,angle=0,width=\FIGUREWIDTH}
}
\vskip 0.3truecm
\caption{Plot of the specific heat $C$ as a function of $\beta$, 
for the same values of $N$ as in Fig.~\ref{fig:03}.
Note that each curve displays a maximum that increases as  $N$ increases.
Since we expect the transition to be asymptotically first-order 
($\phi=1,c>2$), the peaks should develop into a $\delta$-like singularity
as $N\to\infty$.}
\label{fig:04}
\end{figure}

Indeed from Fig.~\ref{fig:03} we note that,
for $\beta<\beta_c\approx 0.9$, $\langle H \rangle/N$ goes 
to zero as $N\to \infty$ whereas $\langle H \rangle/N$ 
reaches a finite limit for $\beta>\beta_c$. 
The crossover between these regimes is particularly sharp and this
is confirmed by the behavior of the specific heat
\beq
C(N,\beta) \equiv 
-\frac{\beta^2}{N}\frac{ \partial \langle H \rangle }{ \partial \beta} =
\frac{\beta^2}{N} (\langle H^2 \rangle - \langle H \rangle^2)
\eeq 
as a function of $\beta$, which is plotted in Fig.~\ref{fig:04}.
In the proximity of $\beta_c$ the singular part of the
specific heat is expected to scale as 
\beq
C \simeq N^{2 \phi-1} f({(\beta-\beta_c)} N^\phi) \ ,
\label{C}
\eeq
 where $f$ is a scaling function and $\phi$ is the crossover exponent.
Since for large $N$ 
one has $C_{\rm max}(N) = \max_\beta C(N,\beta)\sim N^{2\phi-1}$,
by a linear fit of $\ln C_{\rm max}$ vs $\ln N$ we  obtain $\phi =0.9(1)$.
From Eq.~(\ref{C}) we expect also ${(\beta-\beta_c)} \sim N^{-\phi}$  and by 
extrapolating $\beta_c(N)$ vs $1/N^\phi$, with $\phi$ 
given by the previous estimate, we obtain $\beta_c=0.905(5)$.

\begin{figure}[!bt]
\centerline{
\psfig{file=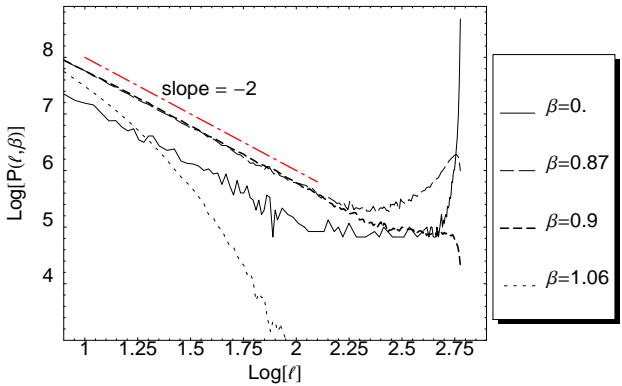,angle=0,width=\FIGUREWIDTH}
}
\vskip 0.3truecm
\caption{A log-log plot of the loop probability distribution
$P(\ell,N=600)$, for different  $\beta$ values.
For $\beta\le 0.9$ the curves display a power law behavior, while
for $\beta= 1.06$ strong cut-off effects are present. }
\label{fig:05}
\end{figure}

As in other recently studied models of DNA denaturation\cite{KMP00,COS02}, 
it is useful to look at the
probability distribution $P(\ell,N)$  of the denatured loops of
 length $\ell$. Indeed, it turns out that for $\beta<\beta_c$  
$P(\ell,N)\sim \ell^{-c}f(l/N)$ 
with an exponent $c$ connected to $\phi$ by the relation
$\phi = \min\{c-1,1\}$, if $c>1$.
Hence, a continuous transition corresponds to $c<2$, while $c>2$
gives $\phi=1$ and a first order denaturation.
To estimate $c$, we examine the $P(\ell, N=600)$, shown in 
Fig.~\ref{fig:05}, for different $\beta$'s.
Its slope at $\beta=0.9$ and in the range $10\le\ell\le 180$
has the borderline value $c=2.01(5)$. This value is slightly lower than 
the one predicted for somewhat simpler models considered recently, 
in which the 
geometry of contacts is drastically simplified to the extent that 
linking entanglement  cannot be defined for the two strands\cite{COS02}. 
This small discrepancy is most probably due to the relatively shorter chains 
considered here and to the less accurate determination of $\beta_c$. We 
believe that the transition has a first order character. This conclusion 
is also supported by the behavior of the linking probability discussed 
below. It should also be taken into account that the present model has 
interactions which induce some bending rigidity of the double stranded 
structure. These effects have already been shown in \cite{COS02} to produce
a slight lowering of the effective $c$ for finite $N$.

\section{Mutual entanglement: linking number}
\label{sec:Lk}
The topological and  geometrical properties of the model are 
then studied by looking at the
behavior, as a function of $T$, respectively 
of the linking probability and of the writhe, which gives a measure
of the degree of supercoiling \cite{BCW80}.  

In this section we ask for the probability that the two strands
are linked as a function of $T$ and, in particular, 
 its behavior at the melting transition, whose location has been 
estimated in the previous section. First let us be clear on what we mean
by two linked curves.

Two disjoint simple closed curves ${{\cal C}_1}$ and ${{\cal C}_2}$ are {\it
topologically unlinked} if there is a homeomorphism of $R^3$ onto itself,
$H: R^3\to R^3$, such that the images $H({{\cal C}_1})$ and 
$H({{\cal C}_2})$ can be separated by a plane\cite{Rolfsen76}.  

A weaker notion of linking is given by the {\it homological linking}
definition.
${{\cal C}_1}$ is {\it
homologically unlinked} from ${{\cal C}_2}$ if ${{\cal C}_1}$ 
bounds an orientable surface
which is disjoint from ${{\cal C}_2}$.  Homological linking is a symmetric
relation, and homological linking implies
topological linking.  In this paper we shall be concerned just with
homological linking. The reasons 
 to restrict ourselves to this definition of linking are twofold. In first 
 place this is the simplest topological property to be checked 
 computationally. In addition, despite the fact that
homological linking is the weakest form of linking, it is known to
be a good indicator of topological linking for configurations in which
the two SAP's are strongly interpenetrating \cite{ORTW94}. 
This is the situation
that we expect to occur in the proximity of the denaturation transition.
 
\begin{figure}[!bt]
\centerline{
\psfig{file=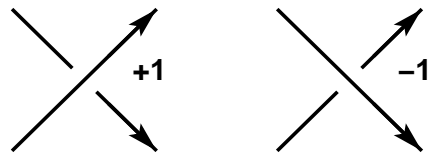,angle=0,width=6cm}
}
\vskip 0.3truecm
\caption{Positive and negative crossings determined a by left-hand rule.}
\label{fig:06}
\end{figure}

A method to detect whether or not the two strands are homologically
linked consists in orienting each of the two strands 
${{\cal C}_1}$ and ${{\cal C}_2}$, and
to project them onto a plane so 
that no vertex in the projection of ${{\cal C}_1}$
coincides with any vertex in the projection of ${{\cal C}_2}$, 
or viceversa (regular projection). 
In this way the mutual crossings are transverse (no edge overlaps 
between the two projections)  and at each point 
where ${{\cal C}_1}$ crosses {\sl under} ${{\cal C}_2}$ we assign a 
value $+1$ or $-1$, according to the orientation of the crossing (see 
Fig~\ref{fig:06}). 
The sum of these crossing numbers is called the {\sl linking number}
of the two curves , $\Lk({{\cal C}_1},{{\cal C}_2})$, 
and the two curves are homologically 
linked if and only if $\Lk({{\cal C}_1},{{\cal C}_2}) \ne 0$\cite{Rolfsen76}.

\begin{figure}[!bt]
\centerline{
\psfig{file=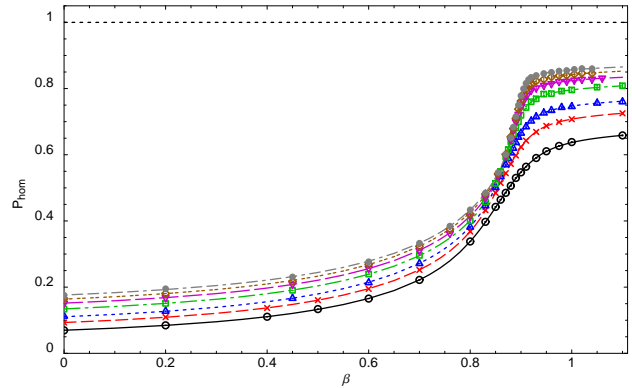,angle=0,width=\FIGUREWIDTH}
}
\vskip 0.3truecm
\caption{Plot of the linking probability as a function of $\beta$. 
Different curves correspond to different $N$  values, as in Fig.~\ref{fig:03}.
In correspondence to the transition ($\beta\approx 0.9$)
an abrupt jump of $P_{\rm hom}$ develops.}
\label{fig:07}
\end{figure}

\begin{figure}[!bt]
\centerline{
\psfig{file=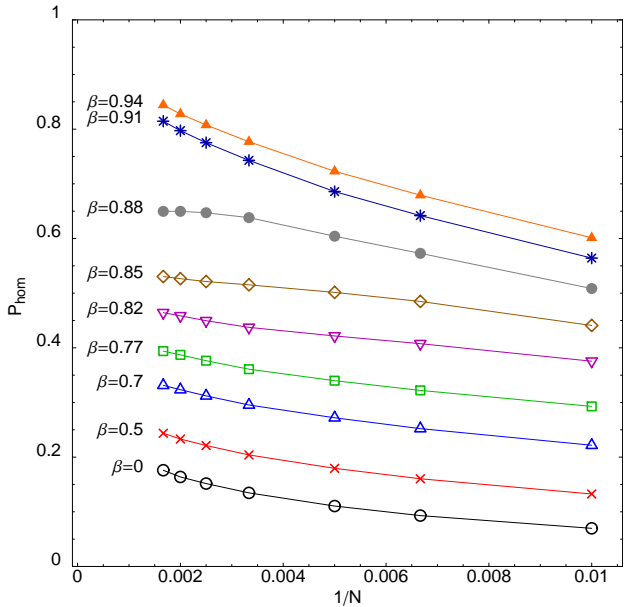,angle=0,width=\FIGUREWIDTH}
}
\vskip 0.3truecm
\caption{Plot of $P_{\rm hom}$ as a function of $1/N$ for different
$\beta$ values.
By observing that $\beta=0.88$ is below the transition point
$\beta_c=0.905(5)$, while  $\beta=0.91$ is just above $\beta_c$,
it is evident that a change in the asymptotic behavior of the curves
takes place in proximity of the transition.}
\label{fig:08}
\end{figure}

\begin{figure}[!bt]
\centerline{
\psfig{file=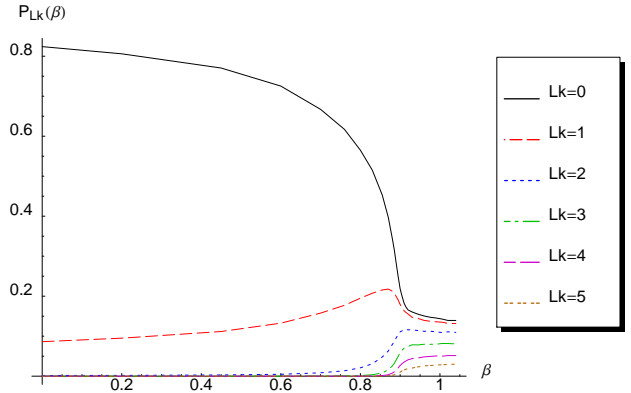,angle=0,width=\FIGUREWIDTH}
}
\vskip 0.8truecm
\caption{Plot of the probability of having a fixed linking number
$\Lk\ge 0$ as a function of $\beta$ for $N=400$.
In the plot we have just considered positive values of $\Lk$ since 
$P_{-\Lk}(\beta) = P_{\Lk}(\beta)$.}
\label{fig:09}
\end{figure}

In Fig.~\ref{fig:07} we show the probability, $P_{\rm hom}$, 
for the two circular 
strands to be homologically linked (i.e. having non-zero linking number) 
as a function of $\beta$ and for different $N$ values. The qualitative
trends are clear. The linking probability increases with increasing 
$\beta$ at fixed $N$. In particular at high $\beta$, (i.e. in the
native phase), the linking probability is very close to $1$, indicating that
the double stranded structures are very likely to be linked {\it even for
finite values of $N$}. It is interesting to notice that,
in the proximity of the denaturation transition, the change in slope is
very sharp, suggesting a possible discontinuity in the value of $P_{\rm hom}$.
This possible feature is more evident in Fig.~\ref{fig:08} where 
$P_{\rm hom}$ has been plotted as a function of $1/N$
for different $\beta$ values. It seems clear that for $\beta<\beta_c$ i.e.
above the denaturation transition, the curves 
extrapolate (as $N\to \infty$) to  values ${\bar{p}} = 
{\bar{p}}(\beta)$ with ${\bar{p}}$ varying gradually, but sensibly, with 
$\beta$,  whereas as soon as $\beta > \beta_c$ (i.e. in the bounded region)
the limiting linking probability is very close to $1$. 
At $\beta=0$ the linking probability seems to extrapolate to a value
${\bar{p}}(0) \approx 0.2 $. 
A more detailed analysis of the linking probability can be carried out
by looking at the probability, $P_{\Lk}(\beta)$,  to have
a given linking number $\Lk$. In Fig.~\ref{fig:09},
$P_{\Lk}(\beta)$ has been plotted
as a function of $\beta$  for different $\Lk$ values and for $N=400$.
Again one can observe that at $\beta \simeq \beta_c$ there is
an abrupt decay of $P_{0}(\beta)$ (i.e. the probability to be unlinked)
to which corresponds a sharp increase in all the
probabilities of having nonzero values of $\Lk$.
For $\beta >>\beta_c$ it seems that $P_0(\beta)$ 
goes to a constant different from zero; 
this is however a finite size effect (the plot is for $N=600$)
and as $N$ increases $P_0(\beta)$ decreases monotonically.

The evidence of a discontinuity of $P_{\rm hom}$ presented above 
supports the first order 
character of the denaturation transition of the model. This should be 
expected in view of the results of Refs.\cite{KMP00,COS02} and of the
fact that $\Lk$ is unrestricted and 
no contributions to the energy are associated here to the twist of the 
bounded DNA segments.

\begin{figure}[!bt]
\centerline{
\psfig{file=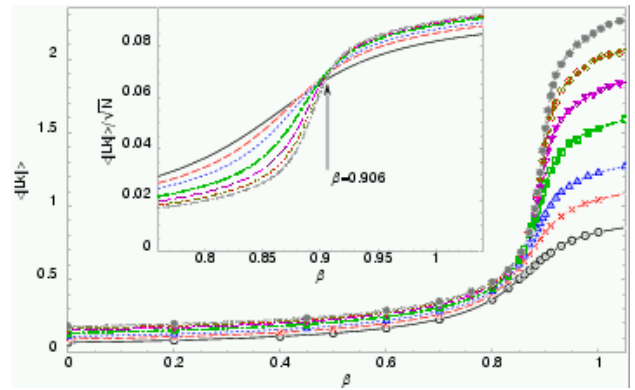,angle=0,width=\FIGUREWIDTH}
}
\vskip 0.3truecm
\caption{
Average of the absolute value of $\Lk$ as a function of $\beta$.
Different curves correspond to different $N$  values, as in Fig.~\ref{fig:03}.
Notice that for $\beta\ge\beta_c$
there is an abrupt increase of the average linking number.
The scaling of $\langle|{\Lk}|\rangle$ is consistent with the form 
$\langle|{\Lk}|\rangle \sim \sqrt{N}$ and with $\beta_c=0.905(5)$, as
shown in the inset.
}
\label{fig:10}
\end{figure}

In Fig.~\ref{fig:10} we show the average of the absolute value of the
linking number $\langle | \Lk | \rangle$ as a function of $\beta$  for 
various values of $N$. In the whole range
of $\beta < \beta_c$ and for each $N$,  $\langle | \Lk | \rangle$ is 
very close to zero and grows very slowly with $N$. 
As $\beta \ge \beta_c$ there is an abrupt change
to a regime characterized by a more rapid 
increase of  $\langle | \Lk |\rangle$ as a function of $N$. 
By assuming that for $\beta > \beta_c$ the following power law behavior
holds
\begin{equation}
\langle | \Lk |\rangle \sim N^{\sigma}
\label{scal}
\end{equation}
we can estimate the $\sigma$ exponent
through  a simple linear fit of $\log \langle | \Lk |\rangle$ vs $\log N$.
This gives
\begin{equation}
\sigma = 0.52 \pm 0.01 \,
\end{equation}
for the whole range $\beta >\beta_c$ (see Fig.~\ref{fig:11}).
It is interesting to notice that such 
estimate is in good agreement with the one obtained 
for the two  boundary curves of an orientable lattice ribbon 
model\cite{ROSTW94}. In other words the low temperature phase
of our model presents the same asymptotic topological properties as those
 of a lattice ribbon
model introduced some time ago to describe the entanglement complexity
of  double-stranded molecules in a good solvent \cite{ROSTW96,ORW96}.
This is what one should expect for a model of dsDNA in the native state
and confirms the adequacy of our model.

A further indication of the scaling behavior
in Eq.~(\ref{scal}) is shown in the inset of Fig.~\ref{fig:10}
where we plot $\langle | \Lk |\rangle /\sqrt{N}$ vs $\beta$: clearly
in the range $\beta > \beta_c$ all the data collapse onto a single
curve whereas in the denatured phase (i.e. $\beta < \beta_c$)
the curve approaches zero as $N$ increases.

\section{Geometrical entanglement: writhe}
\label{sec:Wr}
In this section we analyze the geometrical entanglement of the system
as a function of temperature by computing the writhe of each SAP separately.
In order to define a writhe, consider any simple closed curve in $R^3$, 
and project it
onto $R^2$ in some chosen direction.  In general the projection will have
crossings and, for almost all projection directions, these crossings will
be transverse, so that we can associate a sign $+1$ or $-1$ with each
crossing.  For this projection we perform the sum of these signed crossing
numbers and average over all possible projection directions.  
This average quantity
is the writhe $\Wr$ of the curve \cite{F71}.  The writhe is a geometrical
quantity (since it is not invariant under ambient isotopy) and, contrary 
to the integer $\Lk$,  it is a real
number measuring the extent to which the strand is supercoiled. 
In principle one needs to average the sum of the signed crossing numbers
over all (regular) projections but there is a useful theorem\cite{LacSum91}
 which considerably simplifies the calculation for polygons on the
simple cubic lattice.  
Indeed for a SAP in the cubic lattice, 
the computation of the writhe equals
the average of the linking numbers of the given SAP and its
 push-offs (translate through a sufficiently small
distance) into four non-antipodal octants \cite{LacSum91}.

\begin{figure}[!bt]
\centerline{
\psfig{file=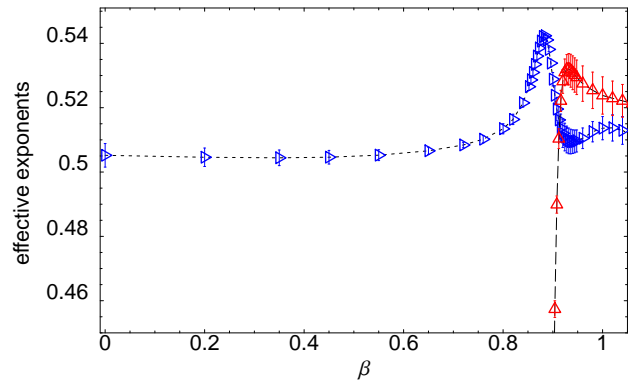,angle=270,width=\FIGUREWIDTH}
}
\vskip 0.3truecm
\caption{Effective exponent $\sigma$ of the linking number ($\triangle$)
and $\zeta$ of the writhe ($\vartriangleright$) as a function of $\beta$.
The variations of  $\zeta$ close to the transition are probably due to
 corrections to the scaling Eq.~(\ref{writhe}) that are much 
larger than the
statistical errors displayed in the figure.
}
\label{fig:11}
\end{figure}

For a single non-interacting SAP it is known that the
mean of the absolute value of the writhe behaves as \cite{ROSTW93}
\begin{equation}
\langle |\Wr| \rangle \sim N^{\zeta},
\label{writhe}
\end{equation}
and we expect the same power law behavior for our model in the denatured
regime (i.e. $\beta < \beta_c$) where the two strands behave essentially
as two independent SAP's.
As a  check we estimate the $\zeta$ exponent at $\beta = 0$ 
assuming the scaling behavior (\ref{writhe}). By a linear fit of the log-log
plot of the $N$ dependence of $\langle |\Wr| \rangle$ at $\beta=0$,
and with $N\in\{300,\,400,\,500,\,600\}$,  we 
obtain
\begin{equation}
\zeta = 0.505(4),
\end{equation}
in full agreement with the corresponding estimate for a single  SAP,
which is also known to be bounded from below by $1/2$\cite{ROSTW93}.
Next, by assuming the validity of (\ref{writhe}) at any $\beta$, we 
estimate the exponent $\zeta$ as a function of 
$\beta$ (see Fig.~\ref{fig:11}). 
With the exception of a little variation in the proximity of the
transition, it seems clear that $\zeta$  remains very close to the
$\beta = 0$ estimate for all $\beta$ values. Thus, the 
denaturation transition does not seem to affect the exponent of the 
power law dependence on $N$ of the absolute value of the writhe. 
This is not surprising if one observes that for $\beta <\beta_c$ the two
circular strands behave essentially as two independent 
SAP's, whereas for $\beta>\beta_c$ the two strands are 
tightly bound together forming a ribbon-like structure.
Indeed the square root of $N$ behavior of the absolute value of the writhe
has been found also for the two boundaries
of an orientable lattice ribbon model\cite{ROSTW94,ROSTW96,ORW96}. 
This confirms, also
from the geometrical point of view, the good correspondence between
the native regime of our model and the lattice ribbon model 
for double stranded 
molecules. The deviations of $\zeta$ from $1/2$
observed in the neighborhood of 
$\beta_c$ are probably due to finite size effects.
However, we can not exclude that a peculiar, distinct scaling regime 
prevails right at the transition. We know that there the statistics of 
denaturated loops is completely different from that at $\beta > \beta_c$ 
and this could imply a modification in the scaling behavior of
$\langle|\Wr | \rangle$. To establish whether a distinct scaling of 
$\langle|\Wr |\rangle$ exists would be, 
however, very challenging  numerically and beyond the scope of 
the present work.

\section{Conclusions}
\label{sec:conc}
In this paper we have studied the topological and geometrical entanglement
of a lattice model for circular double stranded DNA undergoing a denaturation
transition. We have shown that, in the limit of very long chains,  
the linking probability between 
the two strands is a function of $\beta$  in the denaturated 
phase, whereas it jumps
very rapidly to values close to unity as soon as 
$\beta >\beta_c$, i.e. in the bound state.
This feature is confirmed by the behavior of the average linking number
in the two phases: it turns out that  $\langle | \Lk | \rangle$ is a small
constant for $\beta< \beta_c$ and grows roughly as $\sqrt{N}$ in the
low $T$ phase.
This suggests that our model in the native phase is, as far as the 
homological linking is concerned, similar to a ribbon model. This 
analogy is also confirmed by the behavior of the absolute value of the writhe
as a function of $\beta$. In particular $\langle | \Wr| \rangle$
roughly scales
with $N$ as $N^{0.51}$ for all $\beta$'s and no supercoiling effects have been observed
in this model of denaturation.
This can be explained as follows: in our model the linking number is not fixed
to a particular value and can fluctuate freely in the equilibrium statistics. 
Experimentally
this would correspond, at least qualitatively, to the presence in the solution of 
the topoisomerases whose
function is to change the linking number continuously. 
A different scenario can show up
if instead the system is constrained to have a fixed 
linking number (no topoisomerases in solution). 
Indeed the fact that, for our model with unconstrained topology, 
the probability of being linked undergoes, right at denaturation,  
an abrupt (first order like) jump, suggests that the imposition of 
a constraint on the topology of the dsDNA molecule
(by fixing the linking number between the two SAP's in our model) would
affect rather sensibly some features of the transition, such as the melting
temperature. There are indeed experimental indications that supercoiled 
structures characterized by a large (fixed) linking number, display a 
melting transition at higher temperatures than DNA molecules in the 
relaxed state\cite{GBL81}.

Financial support by MURST-COFIN 01, INFM-PAIS 01, and European
Network ERBFMRXCT980183 is gratefully acknowledged.


\end{multicols}


\begin{references}

\bibitem{SKI87} K. Shishido, N. Komiyama and S. Ikawa,
J. Mol. Biol. {\bf 185}, 215 (1987).

\bibitem{DC}
F.B. Dean, A. Stasiak, T. Koller and N.R. Cozzarelli, 
J. Biol. Chem. {\bf 260}, 4795 (1985).

\bibitem{Wang1}
C.D. Lima, J.C. Wang and A. Mondragon, 
Nature {\bf 367}, 138 (1994).

\bibitem{WC}
S.A. Wasserman and N.R. Cozzarelli,
 J. Biol. Chem. {\bf 266}, 20567 (1991).

\bibitem{Wang2}
J. Roca and J.C. Wang, 
 Cell {\bf 77}, 609 (1994).

\bibitem{W69} 
J.H. White,  Am. J. Math.  {\bf 91} 693 (1969).

\bibitem{F71} 
F.B. Fuller,  Proc. Nat. Acad. Sci. U.S.A. {\bf 68}, 815 (1971).

\bibitem{C90} 
N.R. Cozzarelli,
{\sl DNA Topology and Its Biological Effects}
edited by Cold Spring Harbor Laboratory Press (Cold Spring Harbor, NY, 1990).

\bibitem{VC94}
A. V. Vologodskii and N. R. Cozzarelli
Annu. Rev. Biophys. Biomol Struct. {\bf 23}, 609 (1994).

\bibitem{VLRWL65} 
J. Vinograd, J. Lebowitz, R. Radloff, R. Watson, P. Laipis,
Proc. Natl. Acad. Sci. USA {\bf 53}, 1104(1965).


\bibitem{BB78}
R.L. Burke and W. Bauer, Nucleic Acids Res. {\bf 5}, 4819 (1978).

\bibitem{GBL81}
A.~V.~Gagua, B.~N.~Belintev and Y.~L.~Lyubchenko,
Nature {\bf 294}, 662 (1981).


\bibitem{PS66}
P. Poland and H. A. Sheraga, J. Chem. Phys. 45, 1456 (1966).

\bibitem{CCG00}
M. S. Causo, B. Coluzzi, and P. Grassberger,
Phys. Rev. E {\bf 62}, 3958 (2000)

\bibitem{KMP00} 
Y.~Kafri, D.~Mukamel and L.~ Peliti, \prl {\bf 85}, 4988 (2000).

\bibitem{COS02} 
E.~Carlon, E.~Orlandini, and A.~L.~Stella, \prl {\bf 88}, 198101 (2002).

\bibitem{TDP00}
N. Theodorakopoulos, T. Dauxois, and M. Peyrard, 
Phys. Rev. Lett. {\bf 85}, 6 (2000).

\bibitem{CH97}
D. Cule and T. Hwa, Phys. Rev. Lett. {\bf 79}, 2375 (1997).

\bibitem{RB01}
J. Rudnick  and R. Bruinsma, Phys. Rev. E {\bf 65}, 030902(R) (2002).

\bibitem{OBMSM02} 
The model with strictly short range interactions has 
been used for a study of mechanical denaturation of DNA by
E. Orlandini, S. Bhattacharjee, D. Marenduzzo, A. Maritan and F. Seno,
J. Phys. A {\bf 34}, L751, 2001.


\bibitem{TJOW96}
M. C. Tesi,  E. J. Janse van Rensburg , E. Orlandini and S. G. Whittington,
J. Stat. Phys. {\bf 82}, 155 (1996).

\bibitem{MOS90} 
N.~Madras, A.~Orlitsky and L.A.~Shepp, J.~Stat.~Phys. {\bf 58}, 159 (1990).

\bibitem{VS61} 
P.~H.~Verdier and W.~H.~Stockmayer, \jcp {\bf 36}, 227 (1961).

\bibitem{mhm} 
A.~M.~Ferrenberg and R.~H.~Swendsen, \prl {\bf 61}, 2635 (1988).
{\bf 63}, 1195 (1989).


\bibitem{BCW80}
W.R. Bauer, F.H. Crick, and J.H. White, Sci.Am. {\bf 243}, 118 (1980).

\bibitem{Rolfsen76}
Rolfsen D 1976 {\it Knots and Links}, 
edited by Publish or Perish, Inc. (Wilmington)

\bibitem{ORTW94}
 E. Orlandini,E. J. Janse van Rensburg,  M. C. Tesi
and S. G. Whittington,  J. Phys. A {\bf 27}, 335 (1994).

\bibitem{ROSTW94}
E. J. Janse van Rensburg, E. Orlandini, D. W. Sumners, M. C. Tesi
and S. G. Whittington, Phys. Rev. E {\bf 50}, R4279 (1994).

\bibitem{ROSTW96}
E. J. Janse van Rensburg, E. Orlandini, D. W. Sumners, M. C. Tesi
and S. G. Whittington, 
J.  Stat. Phys. {\bf 85}, 103 (1996).

\bibitem{ORW96}
E.\ Orlandini, E.\ J.\ Janse van Rensburg, 
and S. G. Whittington, J. Stat. Phys {\bf 82}, 1159 (1996).

\bibitem{LacSum91} R.C. Lacher and D.W. Sumners
{\it Data Structures and Algorithms for the computation
of topological invariants of entanglements: link, twist and writhe.}  In
Computer Simulations of Polymers, edited by R.J. Roe, ed., Prentice-Hall,
365 (1991).

\bibitem{ROSTW93}
E. J. Janse van Rensburg, E. Orlandini, D. W. Sumners, M. C. Tesi
and S. G. Whittington,  J. Phys. A {\bf 26}, L981 (1993).


\end{references}
\end{document}